\documentclass[aps,pra,twocolumn,superscriptaddress]{revtex4-2}
\usepackage{amsmath,amsfonts,amssymb}
\usepackage{graphicx,float,calc}
\usepackage{color,bm}
\usepackage{ulem}
\usepackage{braket}\hyphenpenalty=5000
\usepackage[colorlinks,urlcolor=blue,citecolor=blue,linkcolor=blue]{hyperref}

\begin{document}
	
\title{Mobility rings in a non-Hermitian non-Abelian quasiperiodic lattice}
	
\author{Rui-Jie Chen}
\affiliation{Key Laboratory of Atomic and Subatomic Structure and Quantum Control (Ministry of Education), Guangdong Basic Research Center of Excellence for Structure and Fundamental Interactions of Matter, South China Normal University, Guangzhou 510006, China}
\affiliation{Guangdong Provincial Key Laboratory of Quantum Engineering and Quantum Materials, School of Physics, South China Normal University, Guangzhou 510006, China}

\author{Guo-Qing Zhang}
\affiliation{Research Center for Quantum Physics, Huzhou University, Huzhou 313000, People's Republic of China}

\author{Zhi Li}
\email{lizphys@m.scnu.edu.cn}
\affiliation{Key Laboratory of Atomic and Subatomic Structure and Quantum Control (Ministry of Education), Guangdong Basic Research Center of Excellence for Structure and Fundamental Interactions of Matter, South China Normal University, Guangzhou 510006, China}
\affiliation{Guangdong Provincial Key Laboratory of Quantum Engineering and Quantum Materials, School of Physics, South China Normal University, Guangzhou 510006, China}

\author{Dan-Wei Zhang}
\email{danweizhang@m.scnu.edu.cn}
\affiliation{Key Laboratory of Atomic and Subatomic Structure and Quantum Control (Ministry of Education), Guangdong Basic Research Center of Excellence for Structure and Fundamental Interactions of Matter, South China Normal University, Guangzhou 510006, China}
\affiliation{Guangdong Provincial Key Laboratory of Quantum Engineering and Quantum Materials, School of Physics, South China Normal University, Guangzhou 510006, China}	
\date{\today}	
	
\begin{abstract}
We study localization and topological properties in spin-1/2 non-reciprocal Aubry-Andr\'{e} chain with SU(2) non-Abelian artificial gauge fields. The results reveal that, different from the Abelian case, mobility rings, will emerge in the non-Abelian case accompanied by the non-Hermitian topological phase transition. As the non-Hermitian extension of mobility edges, such mobility rings separate Anderson localized eigenstates from extended eigenstates in the complex energy plane under the periodic boundary condition. Based on the topological properties, we obtain the exact expression of the mobility rings. Furthermore, the corresponding indicators such as inverse participation rate, normalized
participation ratio, winding number, non-Hermitian spectral structures and wave functions are numerically studied. The numerical results are in good agreement with the analytical expression, which confirms the emergence of mobility rings.
\end{abstract}

\maketitle

\section{\label{sec1}Introduction}
Anderson localization reveals the randomness-induced insulation in disordered systems~\cite{Anderson1958}. The Anderson localization transition is marked by the transition of all extended single-particle states to localized states at a critical disorder strength, which is absent in one and two dimensions under quenched disorders~\cite{Abrahams1979,Lee1985,Evers2008}. However, it can occur in one-dimensional systems with quasiperiodic or correlated disorders. Among various quasiperiodic lattice models, the simplest one is the Aubry-Andr\'{e} (AA) model~\cite{Aubry1980,Roati2008cold,Lahini2009photonic}. The AA model and its generalizations have sparked widespread research interest in many platforms, such as photonic crystals ~\cite{Dal2003photonic}, ultracold atoms ~\cite{Roati2008cold}, and acoustic waves~\cite{Hu2008acoustic}. The original AA model exhibits the Anderson localization transition at a critical quasiperiodic potential strength. In generalized AA models and quasiperiodic systems, the localization transition can be associated with an intermediate phase, with the coexistence of extended and localized eigenstates separated by a critical energy, which is known as the mobility edge~\cite{Sarma1988ME,Sarma2015ME,Sarma2017ME,Immanuel2018ME}. The mobility edges can also emerge in two-dimensional quasiperiodic systems~\cite{Guan2023}.

Recently, considerable effort has been devoted to exploring non-Hermitian physics~\cite{Hatano1996,Bender1998,Ueda2020,Bergholtz2021}. Various exotic properties unique to non-Hermitian systems have been revealed, such as the exceptional points in the complex energy plane~\cite{Bergholtz2021,Ding2022,Sticlet2022,Shen2018}, new topological phenomena ~\cite{Yao2018Aug,Yao2018Sep,Gong2018,Song2019,Jin2019,Yokomizo2019,Xue2020,Yang2020,DW2020TAI,Tang2020,Xue2022TAI}, and the non-Hermitian skin effect~\cite{Yao2018Aug,Jiang2019,li2020critical,DW2020Skinsuperfluid,Zhang2021,LLi2020,zhang2022universal,Longhi2022selfhealing,Zhang2023,Kawabata2023,lin2023topological,SZLi2024NHSE}. Notably, complex energies are common in open systems, whose imaginary parts stand for leakage. The imaginary part of the self-energy is an important notion in the Thouless criterion for Anderson localization~\cite{Thouless1974}. The non-Hermitian skin effect is characterized by the accumulation of all bulk eigenstates near the boundaries, which is associated with the spectral winding topology and modifies the bulk-boundary correspondence in non-Hermitian systems~\cite{Yao2018Aug}.
There are two typical kinds of non-Hermitian Hamiltonians, where the non-Hermiticities come from either the gain-and-loss or the nonreciprocality. The nonreciprocality in lattice systems usually denotes nonreciprocal hopping with asymmetric amplitudes between two lattice sites. Interestingly, some non-Hermitian generalized AA models are studied~\cite{Longhi2019NHquasi,Jiang2019,Tang2021,Shanzhong2024,ChenShu2024,Ren2024}. It has been theoretically found that the Anderson localization transition of eigenstates coincides with the transition of the spectral topology with respect to eigenenergies~\cite{Longhi2019NHquasi,Jiang2019}. The association of the Anderson localization transition and the spectral real-complex transition is broken when some localized modes take complex energies in non-Hermitian AA models~\cite{Wang2025}.

Recent works show that mobility edges can be related to complex eigenenergies in non-Hermitian mosaic quasiperiodic lattices~\cite{Shanzhong2024,ChenShu2024}. Such mobility edges form ring structures in the complex energy plane, which is termed as mobility ring and can be analytically solved based on the Avila's global theory~\cite{Avila2015}. Meanwhile, synthetic gauge fields \cite{Dalibard2011,Goldman2014,Goldman2016,DW2018,Cooper2019} in non-Hermitian systems have attracted increasing attentions in the Abelian case \cite{Longhi2015,Longhi2017bidirection,Longhi2017nonadiabatic,Midya2018,Wong2021imaginarygauge,Lu2021,Shao2022,Yan2020,Yan2022,Wu2022,Li2023enhance,zhang2024skin,Li2024NHAB,RJChen2024,YLi2025}. The interplay of non-Abelian gauge fields and non-Hermiticities remains largely unexplored~\cite{Ezawa2021nAnH,Liang2024nAnH,Zhao2021nAnH,Guo2023,Shan2024nAnH,Zhou2025TheNG}.
Remarkably, a generalized one-dimensional Hatano-Nelson model~\cite{Hatano1996} with imbalanced non-Abelian hoppings is recently proposed~\cite{Pang2024}. Despite lacking gauge flux in one dimension, non-Abelian gauge fields have been shown to induce rich topological consequences, such as the non-Hermitian skin modes at both ends of an open chain. A key problem in this direction is the effect of non-Abelian gauge fields on the Anderson localization transition and mobility rings in non-Hermitian systems.

In this work, we study the localization and topological properties in a spin-1/2 nonreciprocal AA model with imbalanced non-Abelian hopping. In the absence of the non-Abelian gauge fields, the Anderson localization transition under quasiperiodic potential coincides with the real-complex transition in energy spectra under the periodic boundary condition (PBC). In the presence of non-Abelian gauge fields, the Anderson localization transition occurs without accompanying real-complex energy transition. Moreover, we find the topology-dependent mobility edges induced by the non-Abelian gauge field, which are discontinuous in the complex energy plane and absent in the Abelian case. We identify the critical values of quasiperiodic strengths for complex mobility edges under both periodic and open boundary conditions. Combining the non-Bloch approach and the unique topological properties induced by the non-Abelian gauge field, we propose a method to calculate the mobility edges. Our findings reveal that the mobility edges possess a ring-structure, demonstrating the existence of mobility rings in the system. We further reveal the coexistence of Anderson localized modes and two kinds of skin modes under the open boundary condition (OBC) due to the interplay among disorders, non-Abelian gauge fields and non-Hermitian skin effects.

The rest of this paper is organized as follows. In Sec.~\ref{sec2}, we introduce the nonreciprocal AA model with non-Abelian gauge fields. Section \ref{sec3} is devoted to exploring the localization properties with topology-dependent mobility rings in the system. Finally, a short conclusion is given in Sec.~\ref{sec4}.

\section{\label{sec2}Model}

We begin by considering a one-dimensional quasiperiodic AA lattice of spinful particles with nonreciprocal hopping and synthetic non-Abelian gauge fields~\cite{Pang2024}, as shown in Fig.~\ref{fig1}(a). The tight-binding Hamiltonian of the considered system can be written as
\begin{equation}\label{Ham} \hat{H}=\sum_{j}^{L-1}(J_l\hat{c}_j^{\dagger}e^{i\theta_l\hat{\sigma}_y}\hat{c}_{j+1}+J_r\hat{c}_{j+1}^{\dagger}e^{i\theta_r\hat{\sigma}_x}\hat{c}_{j})+\sum_{j}^{L}V_j\hat{c}_j^{\dagger}\hat{c}_j,
\end{equation}
where $\hat{\sigma}_x$ and $\hat{\sigma}_y$ are Pauli matrices for the spin-1/2 particles, $\hat{c}_j^{\dagger}$ ($\hat{c}_j$) is the creation (annihilation) operator on site $j$, and $L$ is the lattice size. Here $J_{l}=Je^g$ and $J_{r}=Je^{-g}$ denote the left- and right-hopping amplitudes with the nonreciprocity parameter $g$, respectively. $V_j=V\cos(2\pi\alpha j)$ is an on-site quasiperiodic potential with the strength $V$, where the quasiperiodicity is achieved by an irrational modulation $\alpha$. When the hopping between sites $j=1$ and $j=L$ is turned on and off, the system is under the PBC and OBC, respectively. Hereafter we set $J=1$ as the energy unit, and $\alpha=(\sqrt{5}-1)/2$ as the inverse golden ratio without loss of generality.

This model features leftward and rightward hopping phases $\theta_l$ and $\theta_r$ with $\hat{\sigma}_y$ and $\hat{\sigma}_x$ spin flips, respectively. Although a gauge flux is not well-defined in a one-dimensional bulk, the Abelian or non-Abelian nature of the associated gauge field can be characterized by the commutation relation of two operators for the left-right hopping $\hat{W}_1=J_l J_re^{i\theta_l\hat{\sigma}_y}e^{i\theta_r\hat{\sigma}_x}$ and the right-left hopping $\hat{W}_2=J_lJ_re^{i\theta_r\hat{\sigma}_x}e^{i\theta_l\hat{\sigma}_y}$ \cite{Yang2019,Pang2024}:
\begin{equation}
	[\hat{W}_1,\hat{W}_2]=J_l J_r \sin\theta_l\sin\theta_r[\hat{\sigma}_x,\hat{\sigma}_y].
\end{equation}
The gauge field is non-Abelian if $[\hat{W}_1,\hat{W}_2]\neq 0$ and Abelian if $[\hat{W}_1,\hat{W}_2]=0$. Thus, the gauge field is non-Abelian only when both $\sin\theta_l$ and $\sin\theta_r$ are non-vanishing. In the absence of the SU(2) non-Abelian gauge field, the Hamiltonian~\ref{Ham} reduces to the nonreciprocal AA lattice \cite{Jiang2019}, where the Anderson localization transition occurs at $V/J=2$ in the Hermitian case of $J_l=J_r$ and is dependent on the nonreciprocal strength with topological nature in the non-Hermitian case of $J_l\neq J_r$, without MEs in both two cases. In the absence of quasiperiodic potentials with $V=0$ in Eq. (\ref{Ham}), the system reduces to the clean Hatano-Nelson model with a synthetic non-Abelian gauge field \cite{Pang2024}. In this situation, the non-Hermitian topology can be modified by such a non-Abelian gauge field, since both the nonreciprocal hopping amplitudes $(J_l,J_r)$ and the non-Abelian hopping phases $(\theta_l,\theta_r)$ contribute to the non-Hermiticity~\cite{Pang2024}. Here we focus on the effect of the non-Abelian gauge field on the Anderson localization transition and related mobility edges in this non-Hermitian quasiperiodic chain.

It is known that mobility edges are identified by the coexistence of extended and localized eigenstates. In non-Hermitian systems, the eigenstate localization transition are associated with spectral winding topology  ~\cite{Longhi2019NHquasi,Jiang2019}. In the following, we take several physical quantities [see Eqs. (\ref{IPR},\ref{NPR},\ref{winding})] to study the localization and topological properties of eigenstates in the system.
For the spin-1/2 particle in the chain, the probability amplitude of the $n$-th eigenstate at the $j$-th site can be written as $\psi_n(j)=(u_n,v_n)^{T}$, 
where $u_n(j)$ and $v_n(j)$ denote the corresponding probability amplitudes of the spin-up and spin-down components, respectively. The localization property of the eigenstate can be characterized by the inverse participation ratio (IPR). The IPR of the $n$-th eigenstate is given by
\begin{equation}\label{IPR}
	\mathrm{IPR}_n=\frac{\sum_{j=1}^{L}\left[\vert u_n (j)\vert^2 + \vert v_n (j)\vert^2\right]^2}{\left[\sum_{j=1}^{L}\vert u_n (j)\vert^2 + \vert v_n(j)\vert^2\right]^2}.
\end{equation}
In the large $L$ limit, one can expect that $\mathrm{IPR}_n\propto L^{-\Gamma}$, where $\Gamma$ is the fractal dimension. For an extended state, $\mathrm{IPR}_n\propto 1/L$ with $\Gamma=1$, whereas $\mathrm{IPR}_n\sim\mathcal{O}(1)$ with $\Gamma=0$ for a localized state. Besides the IPR, one can also use the normalized participation ratio (NPR), which is defined as
\begin{equation}\label{NPR}
	\mathrm{NPR}_n=(\mathrm{IPR}_n\times L)^{-1}.
\end{equation}
In contrast to the IPR, the NPR grows linearly with the system size $L$ for an extended eigenstate while it vanishes for a localized one in the thermodynamic limit. We further use the mean IPR and mean NPR to characterize the localization phase in the system, which are given by $\mathrm{MIPR}=\sum_{n=1}^{D}\mathrm{IPR}_n$ and $\mathrm{MNPR}=\sum_{n=1}^{D}\mathrm{NPR}_n$ averaged over all the eigenstates with $D=L$ being the total number of eigenstates. In the absence of mobility edges, one has $\mathrm{MIPR}\approx0$ and $\mathrm{MNPR}\neq0$ for the extended phase, while $\mathrm{MIPR}\neq0$ and $\mathrm{MNPR}\approx0$ for the localized phase. In the intermediate phase with mobility edges where both extended and localized eigenstates coexist, one has $\mathrm{MIPR}\neq0$ and $\mathrm{MNPR}\neq0$.

Furthermore, we use the winding number of eigenenergies in the complex energy plane to reveal the spectral topology in the non-Hermitian AA chain~\cite{Gong2018,Longhi2019NHquasi}:
\begin{equation}\label{winding}
	w=\int_{0}^{2\pi}\frac{d\phi}{2\pi i} \partial_{\phi} \mathrm{ln} \ \mathrm{det}[\hat{H}(\phi)-E_B],
\end{equation}
under the twisted PBC with the twist phase $\phi\in[0,2\pi]$. Here $\hat{H}(\phi)$ denotes the $\phi$-dependent Hamiltonian given by setting $J_{l}$ ($J_{r}$) as $J_{l}e^{+i\phi}$ ($J_{r} e^{-i\phi}$) in the model Hamiltonian in Eq.~(\ref{Ham}). $E_B$ is an energy base that does not belong to the energy spectrum on the complex plane under the PBC. $w$ counts the number of times the spectrum of $H(\phi)$ winds around $E_B$ when $\phi$ is swept over a period, which is preserved for any $E_B$ within the closed loop. A nonzero spectral winding number $w$ indicates the presence of the non-Hermitian skin effect~\cite{Gong2018,Yao2018Aug,Yao2018Sep}, with $|w|$ localized edge modes of energy $E=E_B$ under the OBC and the sign of $w$ determining the left or right boundary for the skin modes. The non-Hermitian topological transition with respect to the eigenenergies is characterized by the change of the winding number $w$.

\section{\label{sec3}Emergence of mobility rings}

\begin{figure}[t!]
	\centering
	\includegraphics[width=0.47\textwidth]{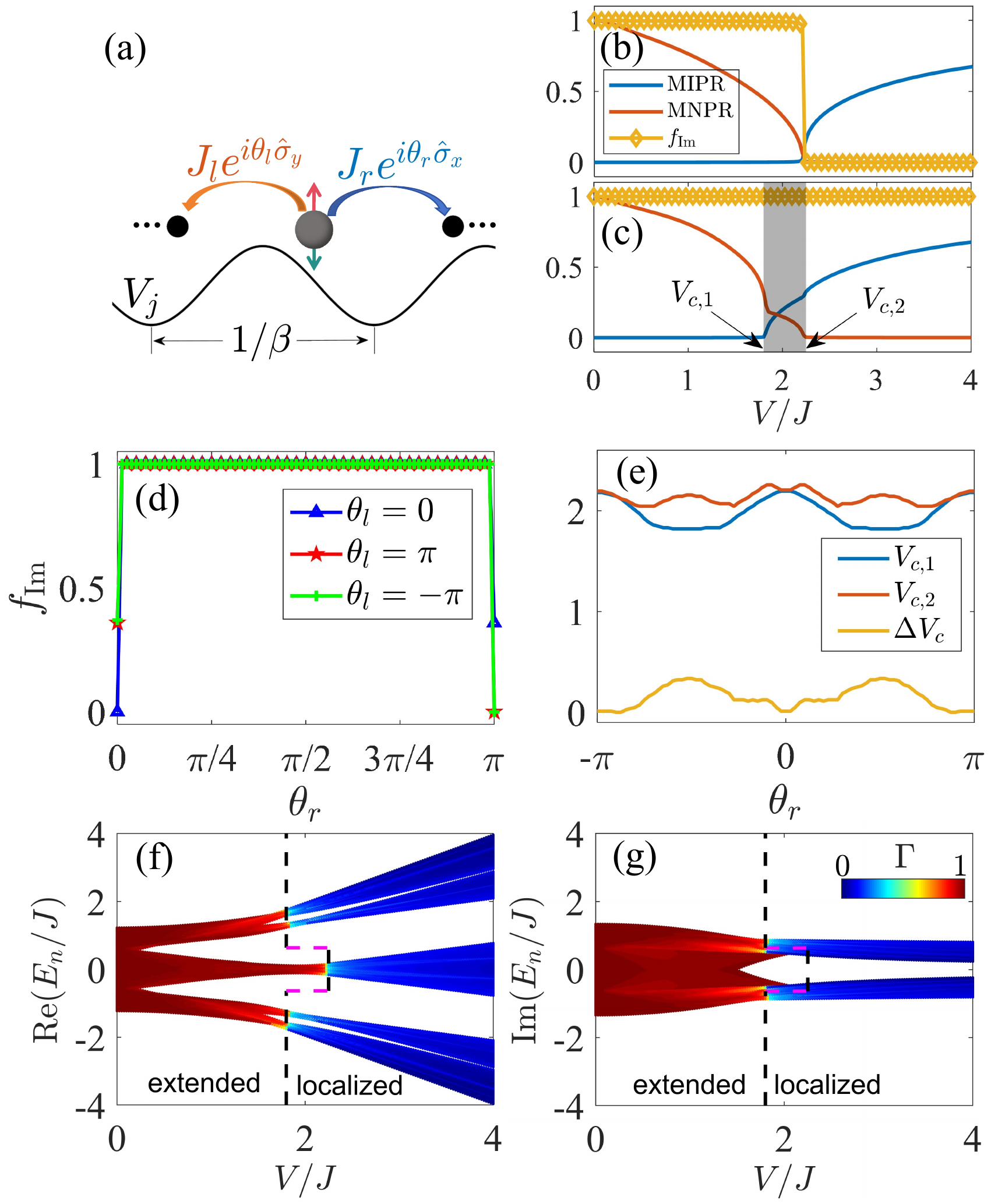}
	\caption{(Color online) (a) Sketch of the nonreciprocal AA chain of spin-$1/2$ particles under synthetic non-Abelian gauge fields. The parameters in (b-g) are $J=1$, $g=0.1$, and $L=987$. (b) MIPR, MNPR and $f_{\mathrm{Im}}$ as a function of $V/J$. The Abelian case with $\theta_l=\theta_r=0$ and the non-Abelian case with $\theta_l=-2.5$ and $\theta_r=-1.4$ are shown in (b) and (c), respectively. The gray shade denotes the intermediate phase with mobility edges. (c) $f_{\mathrm{Im}}$ as a function of $\theta_r$ for $V/J=4$ and various $\theta_l$. (d) Critical points $V_{c,1}$, $V_{c,2}$ and their difference $\Delta V_c$ as a function of $\theta_r$ for $\theta_l=-2.5$. (e) Real part and (f) imaginary part of the energy spectrum superimposed with $\Gamma$. The black and magenta dash lines (see Fig.~\ref{fig3}) denote the complex mobility edges separating the extended and localized eigenstates.}
	\label{fig1}
\end{figure}

Firstly, we compare the localization properties in the nonreciprocal AA model with and without the synthetic non-Abelian gauge field. Figure~\ref{fig1}(b) displays the numerical results of the MIPR and MNPR, and the fraction $f_{\mathrm{Im}}=D_{\mathrm{Im}}/D$ as the ratio of the complex eigenenergies in the whole spectrum. Here $D_{\mathrm{Im}}$ denotes the number of eigenenergies whose imaginary part $|\mathrm{Im}(E_n)|>C$, with the cutoff $C=10^{-13}$ in numerical exact diagonalizations. It can be seen that in the case of $\theta_l=\theta_r=0$, the MIPR and MNPR indicate the Anderson localization transition as a function of $V/J$ without MEs. Moreover, the whole energy spectrum is either real ($f_{\mathrm{Im}}=0$) or complex ($f_{\mathrm{Im}}=1$), and the real-complex transition is accompanied by the Anderson localization transition. In Fig.~\ref{fig1}(d), we show $f_{\mathrm{Im}}$ as a function of $\theta_r$ for fixed $V/J=4$ and various $\theta_l$ to further confirm the absence of real-complex transition under the spin degree of freedom. In the non-Abelian case, the localization property of the system is significantly changed. As shown in Fig.~\ref{fig1}(c), the non-zero values of MIPR and MNPR for a window of the parameter region of $V/J$ imply the presence of the intermediate phase with MEs. We identify the transition point of MIPR and MNPR as $V_{c,1}$ and $V_{c,2}$ and calculate $\Delta V_c\equiv V_{c,2}-V_{c,1}$. Here $V_{c,1}$ and $V_{c,2}$ are chosen as the strength of the on-site potential whose related $\mathrm{MIPR}>C_V$ and $\mathrm{MIPR}<C_V$ with the cutoff $C_V=10^{-3}$ . The extracted $V_{c,1}$, $V_{c,2}$ and $\Delta V_c$ versus $\theta_r$ are plotted in Fig.~\ref{fig1}(e), which shows that the intrinsic non-Abelian gauge field ($\theta_{l,r}\neq0,\pm\pi$) induces the MEs in the nonreciprocal AA model.

%%%%%%%%%%%%%%%%%%%%%%%%%%%%%%%%%%%%%%%%%%%%%%%%%
%%%%%%%%%%%%%%%%%%%%%%%%%%%%%%%%%%%%%%%%%%%%%%%%%

To further reveal the emergence of complex MEs, we calculate the real and imaginary parts of the energy spectra with the fractal dimension $\Gamma$ of individual eigenstates in Figs.~\ref{fig1}(f) and \ref{fig1}(g), respectively. The deep red patches indicate that the extended states become Anderson localized for a finite range of the quasiperiodic strength with $1.8<V/J<2.2$. The eigenstates in the middle of the real part of the spectrum become localized for larger $V/J$ than other eigenstates, similar to those in the imaginary part of the spectrum. These results demonstrate the complex mobility edges induced by the non-Abelian gauge field in the nonreciprocal AA model, where the extended and localized states are separated by critical complex energies in the spectrum.

\begin{figure}[t!]
	\centering
	\includegraphics[width=0.47\textwidth]{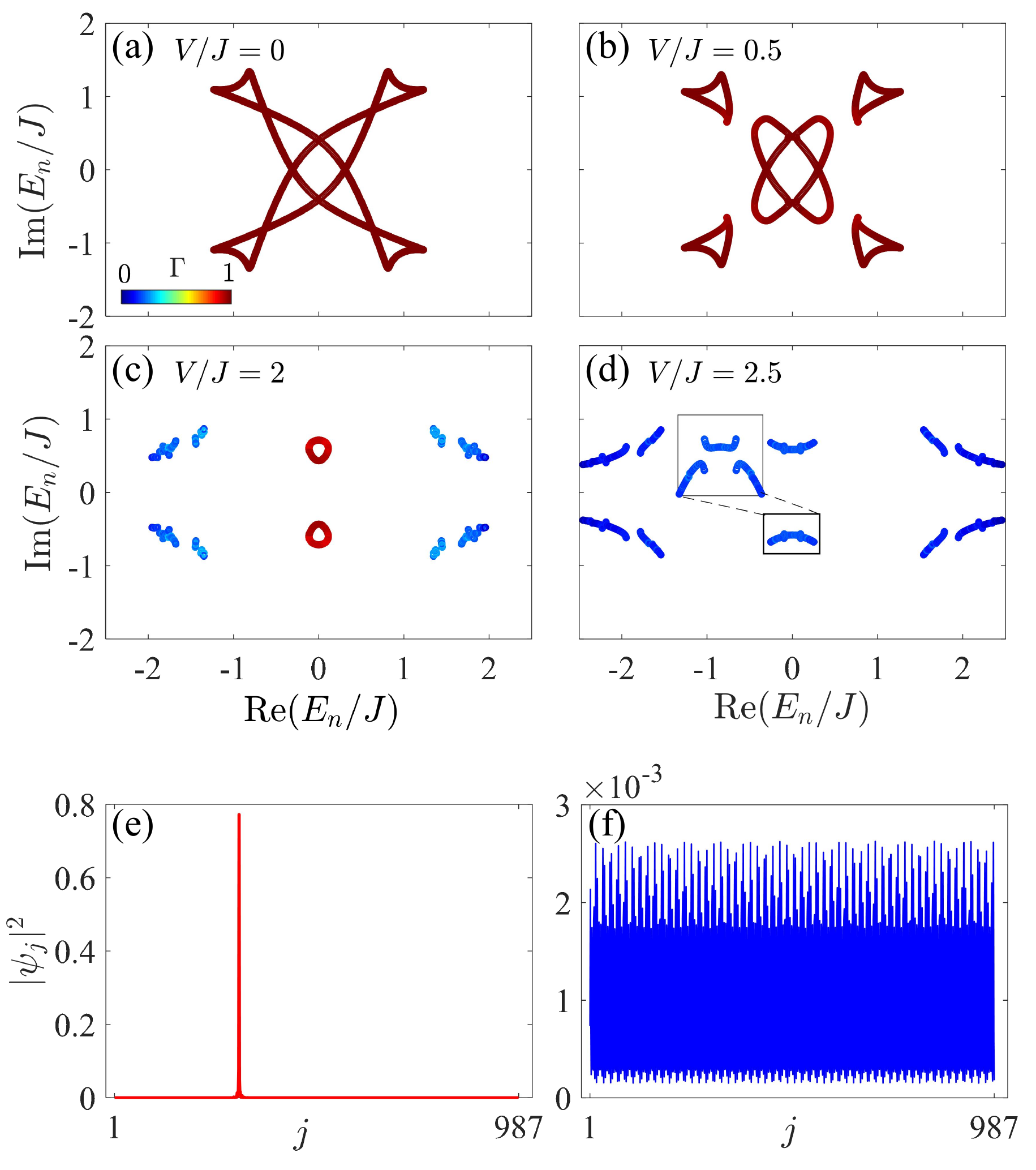}
	\caption{(Color online) Energy spectra and correlative $\Gamma$ under the PBC for (a) $V/J=0$, (b) $V/J=0.5$, (c) $V/J=2$, and (d) $V/J=2.5$, respectively. Real-space distribution of the first eigenstate (e) and the $D/2$-th eigenstate (f) for $V/J=2$. Other parameters are $J=1$, $g=0.1$, $\theta_l=-2.5$, $\theta_r=-1.4$, and $L=987$.}
	\label{fig2}
\end{figure}

Next, we study the non-Hermitian spectral topology related to the Anderson localization transition in the non-Abelian case. We choose a typical set of various quasiperiodic strengths $V/J$ and show the corresponding energy spectra on the complex plane together with the fractal dimension $\Gamma$ of all eigenstates in Figs.~\ref{fig2}(a-d). We also plot the density distributions of the first and the $D/2$-th eigenstates (sorting by the real part of eigenenergies) for $V/J=2$ in Fig.~\ref{fig2}(e) and (f).
In the absence of the quasiperiodic potential with $V/J=0$ in Fig.~\ref{fig2}(a), all eigenstates are extended with $\Gamma\approx1$ and the spectrum forms closed loops in the complex plane under the PBC, indicating the point-gap topology. With the increase of $V/J$ in Fig.~\ref{fig2}(b), the spectrum is separated into several parts, and each part preserves the loop structure and corresponds to extended eigenstates. When $V/J$ is further increased in Fig.~\ref{fig2}(c), some eigenstates with larger values of $|\mathrm{Re}(E_n)|$ (near the four corners of the complex energy plane) become localized with small $\Gamma$ and lose the loop structure. Fig.~\ref{fig2}(e) and (d) shows the density distributions of two localized and extended eigenstates for $V/J=2$. When $V/J$ becomes even larger in Fig.~\ref{fig2}(d), all eigenstates are localized and there is no loop structure in the whole spectrum, indicating the absence of point-gap topology. Thus, we can conclude that the localization transition with complex mobility edges and the non-Hermitian topological transition always happen simultaneously in the nonreciprocal quasiperiodic lattice with the non-Abelian gauge field. When $V/J$ is large enough, the spectrum is fragmented [see Figs.~\ref{fig2}(c) and ~\ref{fig2}(d)], and the complex mobility edges cannot be analytically obtained~\cite{Avila2015}. However, we can numerically extract the complex mobility edges from two critical values of the quasiperiodic strength and the special topological properties induced by the non-Abelian gauge field.

%%%%%%%%%%%%%%%%%%%%%%%%%%%%%%%%%%%%%%%%%%%%%%%%%
\begin{figure}[t!]
	\centering
	\includegraphics[width=0.47\textwidth]{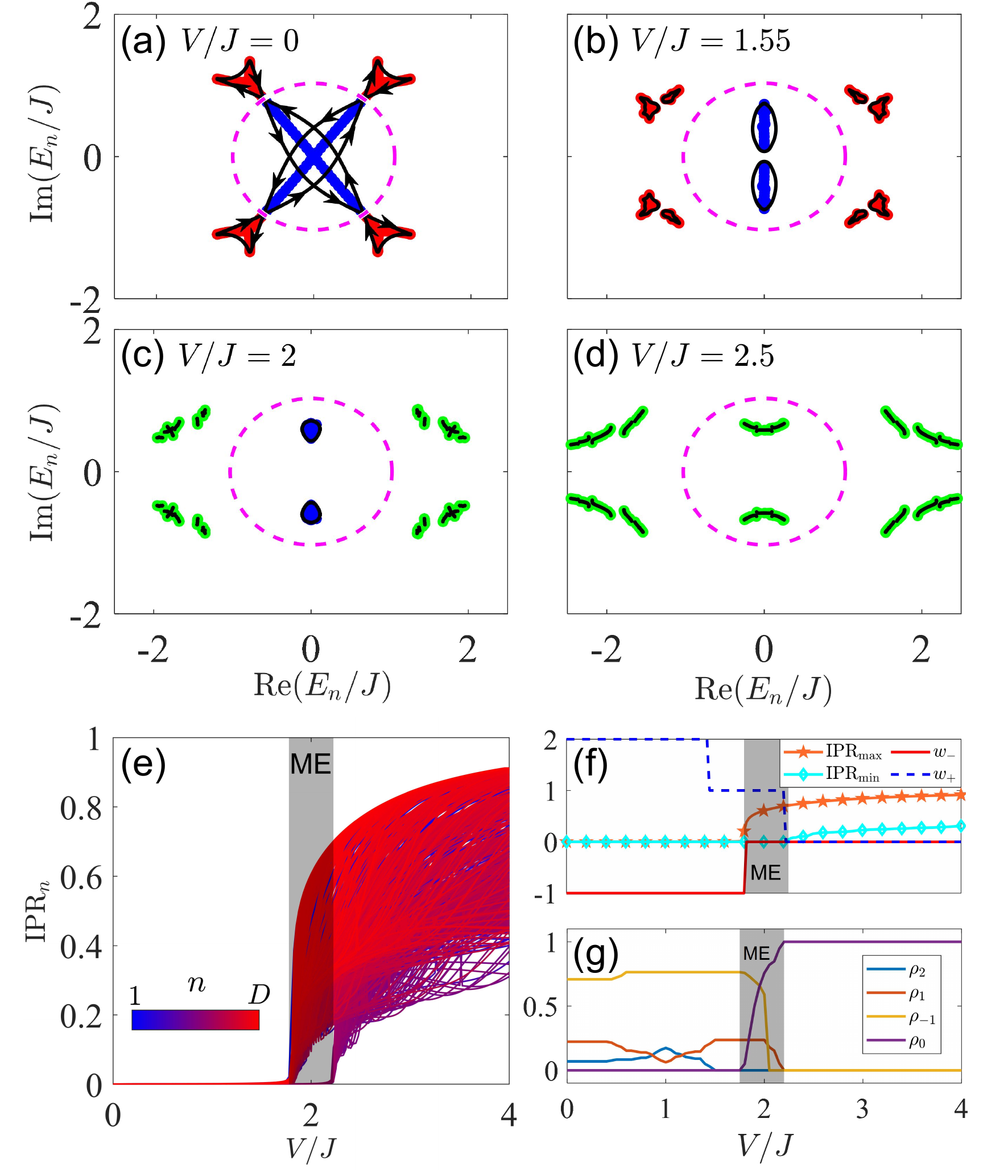}
	\caption{(Color online) Energy spectra under the PBC (black lines) and the OBC (other colors) for (a) $V/J=0$, (b) $V/J=1.55$, (c) $V/J=2$, and (d) $V/J=2.5$, respectively. The red, blue, and green dots (that are too close to appear as lines) correspond to $w<0$, $w>0$, and $w=0$, respectively. The unit circle (magenta dashed line) is obtained from Eq.~(\ref{circle}) and intersects the dividing line of the eigenstates with $w<0$ and $w>0$ under $V/J=0$.
	(e) IPR of all eigenstates as a function of $V/J$.
	(f) Maximal (orange stars) and minimal (cyan diamonds) values of the IPR, the winding numbers $w_{\pm}$ (blue dash line and red solid line) as a function of $V/J$, respectively.
	(g) Proportion of eigenstates $\rho_{0,\pm1,2}$ as a function of $V/J$. The gray shades in (e)-(g) denote the complex mobility edge (denoted by ME). Other parameters are $J=1$, $g=0.1$, $\theta_l=-2.5$, $\theta_r=-1.4$, and $L=987$.}
	\label{fig3}
\end{figure}
%%%%%%%%%%%%%%%%%%%%%%%%%%%%%%%%%%%%%%%%%%%%%%%%%

To further reveal the topology of the mobility edges, we compare the spectra under the PBC and OBC for various $V/J$ in Figs.~\ref{fig3}(a-d). Without quasiperiodic disorders, the energy spectrum under the PBC simultaneously exhibits the clockwise and counterclockwise winding, which are separated by two sectors. The energies under the OBC at the four corners and the center of the spectrum are enclosed by the clockwise-winding sector and the counterclockwise-winding sector, respectively. Thus, we can identify a circle crossing the four intersections of the clockwise-winding and counterclockwise-winding sectors.
The four intersections, while not matching any eigenstates in a finite lattice, correspond to eight extended states in the thermodynamic limit~\cite{Pang2024}. Therefore,  we conjecture that the circle's radius is the modulus of extended eigenenergies. We leverage the non-Bloch approach~\cite{Yao2018Aug,Yokomizo2019,Yang2020} to calculate the energies of extended states. The eigenvalue equation is given by
	\begin{equation}
		\mathrm{det}[\hat{H}(\beta)-E]=0,
	\end{equation}
with $E$ being the eigenenergy. Here $\beta=e^{ik}$ is defined as a phase factor related to complex momentum $k$ for the generalized Brillouin zone of generic non-Hermitian Bloch Hamiltonians~\cite{Yao2018Aug}. The solutions $\beta_i$ can be ordered by their absolute value $|\beta_1|\leq|\beta_2|\leq|...\leq|\beta_{2M}|$ where $2M$ denotes the degree of the equation. The generalized Brillouin
zone $C_{\beta}$ is given by the trajectory of $\beta_M$ and $\beta_{M+1}$ under the condition $|\beta_M|=|\beta_{M+1}|$. In the Hermitian limit, $C_\beta$ is a unit circle. In non-Hermitian cases, the portions of $C_\beta$ inside and outside the unit circle correspond to left and right skin modes~\cite{Yao2018Aug}, respectively.
Now we can pick the complex variable $\beta$ from $C_{\beta}$ to calculate the eigenenergy
	\begin{equation}
		\begin{split}
			E(\beta)=&J_l\mathrm{cos}(\theta_l)\beta+J_r\mathrm{cos}(\theta_r)\beta^{-1}\\&
			\pm i\sqrt{J_l^2\mathrm{sin}^2(\theta_l)\beta^2+J_r^2\mathrm{sin}^2(\theta_r)\beta^{-2}}.
		\end{split}
	\end{equation}
Note that corresponding eigenstate is extended if $|\beta|=1$. Thus, we obtain the equation of the target circle
	\begin{equation}\label{circle}
		x^2+y^2=E(\beta_c)^2
\end{equation}
for $|\beta_c|=1$, where $x$ and $y$ are real numbers and $x+iy$ is in the complex plane. This circle indicates that the mobility edges form a ring structure, which implies the presence of mobility rings~\cite{Shanzhong2024}.

%%%%%%%%%%%%%%%%%%%%%%%%%%%%%%%%%%%%%%%%%%%%%%%%%
\begin{figure}[t!]
	\centering
	\includegraphics[width=0.47\textwidth]{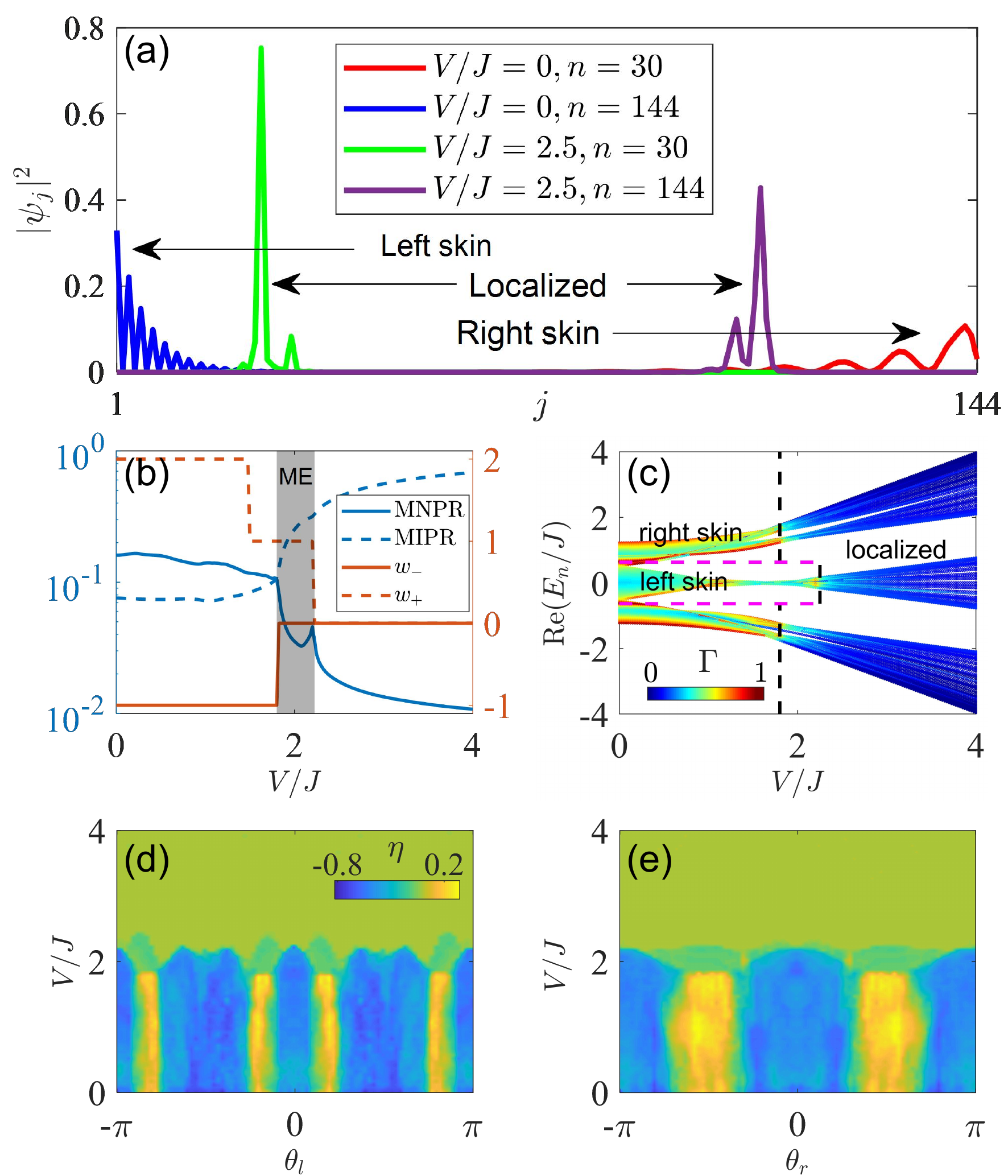}
	\caption{(Color online)(a) Density distributions of left and right skin modes and Anderson-localized modes under the OBC for the $n$-th eigenstate. (b) MNPR, MIPR and winding numbers as a function of $V/J$ for different eigenstates under the OBC. The gray shade denotes the region of mobility edge (denoted by ME) under the PBC. (c) Real part of the OBC spectrum superimposed with $\Gamma$. The black and magenta dashed lines distinguish three kinds of eigenstates under the OBC: the left skin modes, right skin modes, and localized states. $\eta$ as a function of $V/J$ and (d) $\theta_l$ with $\theta_r=-1.4$. (e) Population contrast $\eta$ as a function of $V/J$ and $\theta_r$ with $\theta_l=-2.5$. Other parameters are $J=1$, $g=0.1$, $\theta_l=-2.5$, $\theta_r=-1.4$, and $L=987$.}
	\label{fig4}
\end{figure}
%%%%%%%%%%%%%%%%%%%%%%%%%%%%%%%%%%%%%%%%%%%%%%%%%

Next, we numerically prove that the mobility rings can be used to distinguish the extended and localized eigenstates in the bulk. With the increase of $V/J$, both the two sectors are still separated by the circle. For larger $V/J$, the sector outside the circle does not take the loop structure with $w=0$, while the sector inside the circle still holds the loop-structure with $w\neq 0$. Since the winding numbers $w=0$ and $w\neq 0$ respectively correspond to the localized and extended eigenstates, the circle separates two kinds of eigenstates with different localization properties and critical quasiperiodic strengths.
We compute $\mathrm{IPR}_n$ of each eigenstate in Fig.~\ref{fig3}(e), which shows the range of the mobility ring lying between $V/J\approx1.8$ and $V/J\approx2.2$.
We consider the maxima and minima of the IPRs of all eigenstates as
$\mathrm{IPR_{max}}=\max_{n\in [1,D]} \mathrm{IPR}_n$ and $\mathrm{IPR_{min}}=\min_{n\in [1,D]} \mathrm{IPR}_n$, respectively. Figure~\ref{fig3}(f) displays the results of $\mathrm{IPR_{max}}$, $\mathrm{IPR_{min}}$, and the winding numbers of the energy bases inside and outside the circle denoted as $w_{+}$ and $w_{-}$, respectively. Clearly, $w_{-}$ and $\mathrm{IPR_{max}}$ simultaneously indicate a transition at $V/J\approx 1.8$, while $w_{+}$ and $\mathrm{IPR_{min}}$ indicate another transition at $V/J\approx 2.2$. Notably, $w_{-}$ also shows a transition from a nontrivial $w_{-}=-2$ to $w_{-}=-1$ at $V/J\approx 1.6$ without the Anderson localization transition. We further calculate the proportions of eigenstates with different winding numbers
\begin{equation}
	\rho_w=N_w/D,
\end{equation}
where $N_w$ is the number of eigenstates with the winding number $w=0,\pm1,2$. As shown in Fig.~\ref{fig3}(g), with the increase of $V/J$, $\rho_0$ begins to increase at $V/J\approx1.8$ and approaches nearly one after $V/J\approx 2.2$, while $\rho_{\pm1,2}$ decrease in this intermediate regime and then tend to zero. Thus, we demonstrate and numerically extract mobility rings in this non-Hermitian quasiperiodic chain with non-Abelian gauge fields, such as the black and magenta dashed lines shown in Figs.~\ref{fig1}(f) and \ref{fig1}(g).

\begin{figure}[t!]
	\centering
	\includegraphics[width=0.35\textwidth]{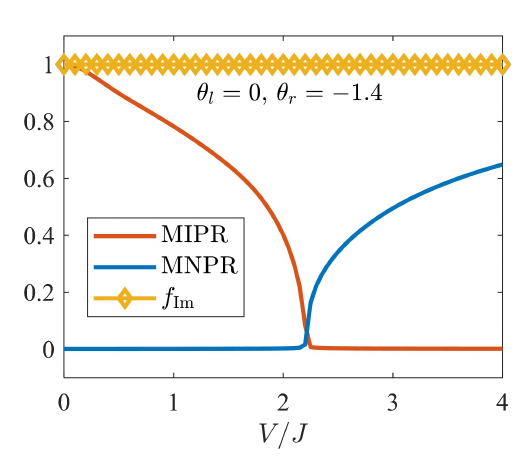}
	\caption{(Color online) MIPR, MNPR and $f_{\mathrm{Im}}$ as a function of $V/J$. Other parameters are $\theta_l=0$, $\theta_r=-1.4$, $J=1$, $g=0.1$, and $L=987$.}
	\label{fig5}
\end{figure}

Finally, we study the localization properties and the non-Hermitian skin effect under the OBC, which are relevant to the mobility rings under the PBC. Two types of winding $w_{\pm}$ in the PBC spectrum correspond to the coexistence of left and right skin modes localized near the two edges of the chain, as shown in Fig.~\ref{fig4}(a). When the quasiperiodic potential is sufficiently strong [such as $V/J=2.5$ in Fig.~\ref{fig4}(a)], the eigenstates become Anderson localized in the bulk. Thus, the eigenstates in the system can undergo a skin-localization transition under the OBC. Moreover, we find that this transition corresponds to the intermediate phase under the PBC. As shown in Fig.~\ref{fig4}(b), the corresponding results of the MIPR, MNPR, and $w_{\pm}$ show the same transition point as the mobility rings determined in the PBC case. Fig.~\ref{fig4}(c) further shows the real part of the OBC spectrum superimposed with the fractal dimension $\Gamma$, where the two dashed lines with different colors denote the region for the mobility rings. Here the boundary between the right-skin and left-skin modes remains unchanged with the increase of $V/J$ and terminates with the critical values of $V/J$. Note that the mobility ring also occurs in the imaginary part of the spectrum, which is similar to that in Fig.~\ref{fig1}(g) and not shown here. We define the population contrast of two types of skin modes $\eta$ as
\begin{equation}
     \eta=(N_r-N_l)/D,
\end{equation}
where $N_l$ and $N_r$ are the numbers of the left and right skin modes, respectively. To show the dependence of the skin localization on the non-Abelian gauge field under the OBC, we plot the results of $\eta$ in the $V$-$\theta_l$ and $V$-$\theta_r$ planes in Figs.~\ref{fig4}(d) and \ref{fig4}(e), respectively. When $V/J=0$, the variation of $\eta$ with $\theta_l$ and $\theta_r$ has been given in Ref.~\cite{Pang2024}. As $V/J$ increases up to $V/J<V_{c,1}$, $\eta$ is almost unchanged as both left-skin and right-skin modes persist before the Anderson localization transition. When $V_{c,1}<V/J<V_{c,2}$, the right-skin modes become Anderson localized modes while the left-skin modes persist [see Fig.~\ref{fig4}(c)], such that the region of $\eta>0$ is reduced. In the region of  $V/J>V_{c,2}$, one has $\eta\approx0$ as both two kinds of skin modes become Anderson localized modes.

\begin{figure}[t!]
	\centering
	\includegraphics[width=0.4\textwidth]{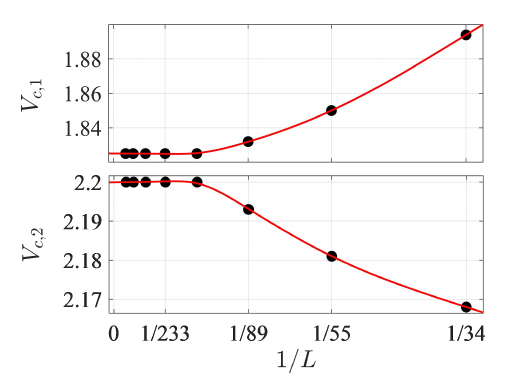}
	\caption{(Color online) Two critical values $V_{c,1}$ (upper) and $V_{c,2}$ (lower) as a function of $1/L$. Other parameters are $\theta_l=-2.5$, $\theta_r=-1.4$, $J=1$ and $g=0.1$.}
	\label{fig6}
\end{figure}

Before concluding, we discuss a intermediate case with $\theta_l=0$ and $\theta_r\neq 0$ or $\pm\pi$. In this case, the system is Abelian. As shown in Fig.~\ref{fig5}, the MIPR and MNPR still undergo transitions simultaneously, while the whole energy spectrum remains entirely complex as $V/J$ is increased. This indicates the absence of mobility ring and the real-complex transition. Based on the conclusions from the previous discussion, we can deduce that the breaking of the real-complex transition is induced by the spin degree, while mobility ring is induced by the non-Abelian gauge field. We also study the finite-size effect. We note that the results for the system size $L=987$ in our numerical simulations are sufficiently large with negligible finite-size effects. To verify this point, for instance, we perform the finite-size scaling of critical values $V_{c,1}$ and $V_{c,2}$ in Fig.~\ref{fig6}. One can find that $V_{c,1}$ and $V_{c,2}$ approach the constant values as $L$ is increased up to $L\gtrsim 144$.

\section{\label{sec4} Conclusion}

In conclusion, we have explored the localization and topological properties of a spinful nonreciprocal AA model with synthetic non-Abelian gauge fields. We have found that the Anderson localization transition generally occurs without accompanying the real-complex energy transition, but still coincides with the transition of the spectral topology. Moreover, the system exhibits topology-dependent mobility rings induced by non-Abelian gauge fields, which are absent in the Abelian case. We obtain the exact expression of mobility rings and further investigated their interplay with the non-Hermitian skin effect. This mobility ring separates the left-skin modes, right-skin modes, and Anderson localized modes under the OBC. Furthermore, the corresponding indicators have been investigated through numerical methods. All the numerical results are in good agreement with analytical ones. Our results reveal intriguing localization and topological phenomena under the interplay among quasiperiodic disorders, non-Abelian gauge fields and non-Hermitian effects.

\begin{acknowledgments}
This work is supported by the National Natural Science Foundation of China (Grants No. 12174126 and No. 12104166), the Guangdong Basic and Applied Basic Research Foundation (Grant No. 2024B1515020018), the Science and Technology Program of Guangzhou (Grant No. 2024A04J3004), and the Open Fund of Key Laboratory of Atomic and Subatomic Structure and Quantum Control (Ministry of Education).
\end{acknowledgments}

\bibliography{reference}

\end{document}